\documentclass[oldversion]{aa}

\usepackage{epsfig}
\usepackage{graphics}
\usepackage{float}
\usepackage{amsmath}
\usepackage{multirow}
\usepackage{longtable}
\usepackage{rotate}
\usepackage{array}
\usepackage{subfigure}
\def\kms{\ifmmode{\rm km\,s^{-1}}\else\hbox{$\rm km\,s^{-1}$}\fi}

\setlongtables

\begin{document}

\title{Analysing weak orbital signals in Gaia data}

\author{L.B.Lucy}
\offprints{L.B.Lucy}
\institute{Astrophysics Group, Blackett Laboratory, Imperial College 
London, Prince Consort Road, London SW7 2AZ, UK}
\date{Received ; Accepted }

\abstract{Anomalous orbits are found when minimum-$\chi^{2}$ estimation
is applied to synthetic {\em Gaia} data for 
orbits with astrometric signatures comparable to the single-scan
measurement error (Pourbaix 2002, A\&A,385,686). These orbits are nearly 
parabolic, edge-on, and their major axes align with the line-of-sight
to the observer. Such orbits violate the 
Copernican principle (CPr) and as such could be rejected.
However, the preferred alternative is to 
develop a statistical technique that  
incorporates the CPr as a fundamental postulate. This can be achieved
in a Bayesian context by defining a Copernican prior.
Pourbaix's anomalous orbits then no longer arise.
Instead, the selected orbits have a somewat higher $\chi^{2}$ but do not 
violate the CPr. The problem of detecting a weak additional orbit in an astrometric binary with a well-determined orbit is also treated. 
}

\authorrunning{Lucy}
\titlerunning{Weak orbits}
\maketitle
%
%
%
%
%
%
%
%
%
%
\section{Introduction}

With the {\em Gaia} observatory in orbit at $L2$
and with commissioning underway,
astronomers can look forward with increasing confidence to the eventual
release of an enormous
quantity of high precision astrometric data.
Initially, this data will be analysed with the already-existing pipeline 
software created by the various consortia. The resulting pipeline products
will no doubt be entirely satisfactory for the vast majority of observed
objects. However, a lesson from earlier large-scale surveys 
is that a small number of objects at the limit of a 
survey's range often prove to be of exceptional interest. For such objects,
standard reduction techniques may give anomalous and misleading results.

This occurred for the Hipparcos mission.
As reviewed by Pourbaix (2004) and Perryman (2009, p.594), 
orbits fitted to Hipparcos data for stars with known spectroscopic orbits 
led to ``discoveries'' that were later refuted. As emphasized by
Pourbaix (2004), ``fitting the noise with an orbital model can have some awful 
consequences''.

This earlier episode suggests that the extraction
of orbital parameters from weak orbital signals in {\em Gaia} data should be
investigated.
In fact,
this is already the subject of an intriguing paper by Pourbaix (2002).
He found that min-$\chi^{2}$ solutions for weak orbits are frequently
anomalous - specifically, edge-on and nearly parabolic. 
In the present paper, the origin of such orbits is explained and a
Bayesian technique developed that
overcomes this problem.
\section{Synthetic data}

In this section, synthetic 1D scans of a model astrometric binary are created.
In order to focus on orbital parameters, we follow
Pourbaix (2002) in assuming that 
parallactic and proper motions have been subtracted.
With regard to notation, previous papers (Lucy 2014a,b; hereafter L14a,b) 
are followed closely.  
\subsection{Orbital elements} 

In contrast to L14a,b, the secondary is here not detected, so the 
astrometry measures the primary's reflex motion about the system's 
centre of mass.   
This motion is parameterized with the Campbell elements 
$P,e,T,a,{\rm i},\omega,\Omega$. 
Here $P$ is the period, $e$ is the eccentricity,
$T$ is a time of periastron passage, ${\rm i}$ is the inclination, 
$\omega$ is the longitude of 
periastron, and $\Omega$ is the position angle of the ascending node.
However, following many earlier investigators
- references in L14a  - the Thiele-Innes elements are also used,
thereby exploiting the resulting linearity in four parameters.
Thus, the Campbell vector $\vec{\theta} \equiv (\vec{\phi},\vec{\vartheta})$, 
where $\vec{\phi} = (P,e,\tau)$, and where 
$\vec{\vartheta} = (a,{\rm i},\omega,\Omega)$ is 
replaced by the vector $\vec{\psi}$ whose
components are the Thiele-Innes constants
$A,B,F,G$.   
(Note that in $\vec{\phi}$, periastron has been replaced by $\tau = T/P$ 
which by definition $\in (0,1)$.)
\subsection{Model astrometric binary} 

The model binary has the following elements:
\begin{eqnarray}
  P_{*}=2.9y &  \;\;\; e_{*}=0.05  &  \;\;\;  \tau_{*}=0.4     
                                                   \nonumber    \\
  \;\;\; a_{*} = \beta \: \sigma &  \;\;\;  
  {\rm i}_{*}=40\degr   &  \;\;\;   \omega_{*} = 150\degr  \;\;\;\;        
                                              \Omega_{*} = 70\degr
\end{eqnarray}
Note that $P_{*}$ is less than $t_{M} = 5y$, the duration of the {\em Gaia} 
mission, so that the issue of incomplete orbits (L14a) is not of concern here.
Also the semi-major axis $a_{*}$ is expressed as a dimensionless multiple 
$\beta$ of the standard error $\sigma$ of a single-scan measurement.
Thus our ability to detect weak orbits can be investigated by
letting $\beta \rightarrow 0$. 

The eccentricity $e_{*}=0.05$ is typical for giant planets in the solar system.
But the main reason for such a small value is to highlight the
anomaly when nearly parabolic orbits fit the data. 

\subsection{Observing campaign} 

A {\em Gaia}-like observing campaign is defined by $t_{n}$, 
the $N$ times 
at which the star is scanned, by $\alpha_{n}$, the corresponding scanning 
angles,
and by $\sigma$. We take $t_{n} = t_{M} z_{u}$  
and $\alpha_{n} = 2\pi  z_{u}$, 
where the $z_{u}$ here and later denote {\em independent}
random numbers $\in (0,1)$. 

Although $\beta = a_{*}/\sigma$ is the important parameter, we take
$\sigma = 40 \mu as$, the expected accuracy for a single transit at 
G-band magnitude $\approx 14$ (see Fig.2 in Sozzetti et al. 2014). 
Note that, in the comprehensive investigation of planet detection with
{\em Gaia} by Casertano et al. (2008), $\sigma = 8 \mu as$.

From Fig.1 in Sozzetti et al. (2014),
we take $N = 70$ as a representive number of scans during the mission. 
\subsection{Synthetic scans} 

Given $\beta$, Eq.(1) defines the theoretical orbit.
The Cartesian sky coordinates $(x^{*}_{n},y^{*}_{n})$ at $t_{n}$
can therefore be computed from Eqns.(A.2) of L14a. The corresponding
1D coordinate or abscissa is 
$s^{*}_{n} = s_{n}(x^{*}_{n},y^{*}_{n})$, where
\begin{equation}
  s_{n} = x_{n} \: cos \: \alpha_{n} + y_{n} \: sin \: \alpha_{n} 
\end{equation}
Here $\alpha_{n}$ is the angle between the scanning direction and 
the $x$-axis - see Fig.1 in Pourbaix(2002). A synthetic data set is 
then
\begin{equation}
  \tilde{s}_{n} = s^{*}_{n} + \sigma \: z_{G}  
\end{equation}
where the $z_{G}$ are independent random Gaussian variates sampling
${\cal N}(0,1)$. Note that the $\chi^{2}$ of the measurement errors is
simply
\begin{equation}
  \tilde{\chi}^{2} = \sum_{n} \: z_{G}^{2} 
\end{equation}

The $N$-dimensional vector $\tilde{\vec{s}}$ with elements $\tilde{s}_{n}$
is the data vector from which orbital elements
are to be estimated. For a given orbit
$\vec{\theta} \equiv (\vec{\phi},\vec{\psi})$, the goodness-of-fit
to $\tilde{\vec{s}}$ is measured by 
\begin{equation}
  \chi^{2} = \frac{1}{\sigma^{2}} \sum_{n} (\tilde{s}_{n} - s_{n})^{2}
\end{equation}
where $s_{n} = s(t_{n},\alpha_{n}; \vec{\theta})$.
\section{Feasible orbits}

In this section, a procedure from L14a is used to explore the likely
degradation of extracted orbits as $\beta \rightarrow 0$.
\subsection{Grid scan} 

A 3-D grid in the $\vec{\phi}$ variables is set up as follows:
the mid-point of grid cell $(i,j,k)$ is $(\log P_{i}, e_{j}, \tau_{k})$.
The grid has $200$ constant steps in each of these variables, with ranges 
$(0.0,1.0)$ for $\log P$, and $(0,1)$ for $e$ and $\tau$.

For specified $\beta$, a synthetic scan  
vector $\tilde{\vec{s}}$
is created as described in Sect.2.4. Then, at each grid point,
the min-$\chi^{2}$ Thiele-Innes vector $\hat{\vec{\psi}}$  
is computed as described in Appendix A.1. The resulting 
$\chi^{2} = \hat{\chi}^{2}_{ijk}$.  
\subsection{Feasible domain ${\cal D}$} 

An orbit $\vec{\theta}_{ijk} = (\vec{\phi}_{ijk},\hat{\vec{\psi}})$ is
deemed to be feasible if
\begin{equation}
  Pr(\chi^{2} > \hat{\chi}^{2}_{ijk}) > 0.05 
\end{equation}
and the ensemble of such orbits
define the feasible domain(s) ${\cal D}$ in $\vec{\phi}$-space.

For $\beta \ga 10$, the domain ${\cal D}$ is a small ellipsoidal volume 
approximately centred on the exact values $(\log P_{*}, e_{*}, \tau_{*})$.
But as $\beta$ decreases, ${\cal D}$ increases and eventually 
develops extraordinary topology.
 
From a sequence of grid scans with $\beta \rightarrow 0$, the value  
$\beta = 2.8$ is found to be such that ${\cal D}$ just extends to $e = 1$. 
Figure 1 illustrates the resulting distortions of 
${\cal D}$.
In this figure, a filled circle is plotted at $(\omega,e)$ 
if Eq.(6) is 
satisfied, and we see that this projection of the feasible orbits
$\vec{\theta}_{ijk}$ extends far beyond the exact values 
$(150\degr, 0.05)$. The most notable features are the 
two narrow
spikes that emerge at $e \sim 0.4$ and reach $e = 1$ at {\em precisely} 
$\omega = 90\degr$ and $270\degr$.  
\begin{figure}
\vspace{8.2cm}
\includegraphics{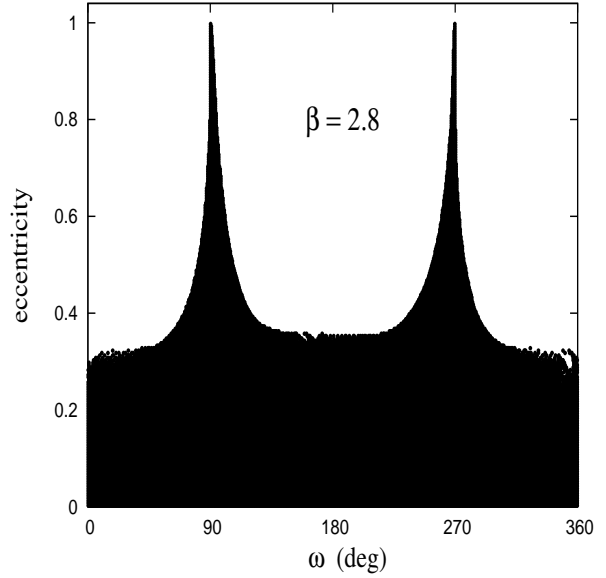}
\caption{Feasible domain ${\cal D}$ projected on to the $(\omega,e)-$plane.
The orbital parameters are given by Eq.(1) with $\beta = 2.8$.} 
\end{figure}
Further information about these spikes is provided by other projections
of the $\vec{\theta}_{ijk}$. In Fig.2, the vectors are projected
onto the  $(\rm i, e)$-plane, and this shows that along 
both spikes $ \rm i \rightarrow 90\degr$ as $e \rightarrow 1$. Accordingly,
if the orbital signal is weak enough $(\beta < 2.8)$, an acceptable fit  
is provided by nearly parabolic, edge-on orbits with $\omega = \pi/2$ or 
$3 \pi/2$.
\begin{figure}
\vspace{8.2cm}
\includegraphics{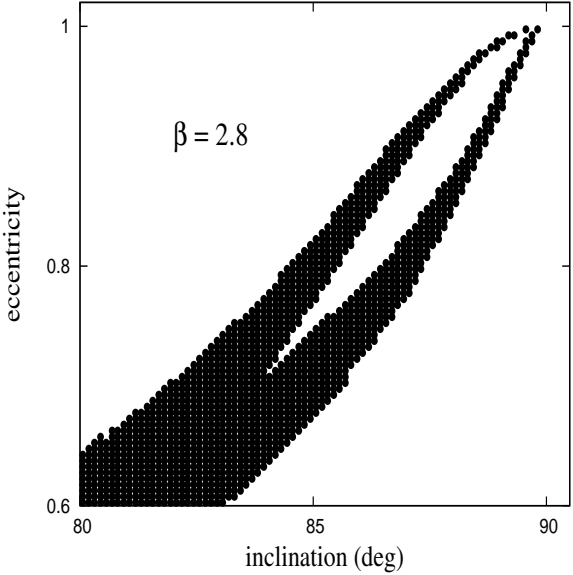}
\caption{Feasible domain ${\cal D}$ projected on to the $(\rm i,e)-$plane.
The orbital parameters are given by Eq.(1) with $\beta = 2.8$.} 
\end{figure}
But this remarkable finding is not original: 
Pourbaix(2002), in reporting least-squares fits to synthetic
1D scans, found an accumulation of nearly parabolic orbits
when $\beta = a_{*}/\sigma = 1.33$ and noted that such orbits lead to 
reasonable {\em apparent} orbits when $\rm i$ and $\omega$ are close to 
$\pi/2$. This serendipitous numerical discovery posed what he called 
``the puzzling case of almost parabolic orbits''. 
\subsection{Violations of the Copernican Principle (CPr)}

When the observed star is at periastron (t = T), its Cartesian
coordinates are:
\begin{eqnarray}
 x = &  a_{p} & \! ( cos \Omega \: cos \omega - sin \Omega \: sin \omega  \: cos \rm i) 
                                              \nonumber    \\
 y = & a_{p} & \! ( sin \Omega \: cos \omega + cos \Omega \: sin \omega  \: cos \rm i) 
                                               \nonumber    \\
 z = & a_{p}  & sin \omega \: sin \rm i
\end{eqnarray}
where the periastron distance $a_{p} = a(1-e)$. Accordingly, if 
$\rm i = \pi/2$ and $\omega = \pi/2$ or $3 \pi/2$, periastron has coordinates 
$(0,0, \pm a_{p})$. 
The major axis is thus aligned
with the line-of-sight to the observer, who therefore finds himself in
a special location. The observer might then object that this orbit violates the 
CPr.   
But this objection could be raised against a slightly non-circular orbit 
with the same $(\rm i, \omega)$,
and an edge-on circular orbit is not particularly
objectionable from the standpoint of the CPr.
To fully appreciate the CPr violation, 
the extra fact that Pourbaix orbits (hereafter P-orbits) are nearly parabolic
must therefore be taken into account.  

Consider the astrometric signal's dependence on orbit orientation. 
Since the maximum elongation of the star from the barycentre is $a(1+e)$,
the extreme range for the abscissae $s_{n}$ is $-a(1+e)$ to $+a(1+e)$.
However, when $\rm i = \pi/2$ and $\omega =  \pi/2$ or $3 \pi/2$, this
range shrinks to its minimum, namely $-b$ to $+b$, 
where $b = a \sqrt{1-e^{2}}$ is the semi-minor axis
Thus, for the Pourbaix solutions, the orbit's {\em inferred} orientation 
{\em and} 
eccentricity are such that the signal is at a deep minimum.
For example, the ratio of the maximum to minimum elongations is 
$\sqrt{(1+e)/(1-e)}= 14.1$ when $e = 0.99$. 
This is a 
large effect, and so the observer would be correct 
in concluding that P-orbits violate the CPr.

Traditionally, when an analysis leads to a CPr 
violation, astronomers suspect that some underlying hypothesis $H$ 
must be wrong. A classic 
example is Herschel's model of the Milky Way, which
violates the CPr because the sun is close to its centre.
In this case, the error is Herschel's implicit assumption that interstellar
space is transparent.

But note a crucial difference.
In these simulations - and in those of L14a - CPr 
violations 
arise even though $H$ - Keplerian motion - is rigorously correct. 
This strongly implies that there must exist a data-analysis technique
that includes the CPr {\em ab initio} rather than invoking it to pass 
judgement on a model only after it has been derived. 
\subsection{Degeneracy}

For a single star, the astrometric
solution has five parameters: the star's right ascension (RA) and declination
at a reference epoch, two components of proper motion, and its parallax.
However, because of errors in the $\tilde{s}_{n}$, this solution 
has residuals, and so it is likely that the addition of orbital motion
(seven parameters) will ``improve'' the fit - i.e., reduce $\chi^{2}$. 
Given their minimal astrometric signatures (Sect.3.3), P-orbits with
$b \la \sigma$ can be added with
little effect on the fit. Evidently, a single-star solution is degenerate
under the addition of a P-orbit with {\em arbitrarily large} semi-major
axis $a$ so long as the semi-minor axis
$b = a \sqrt{1 - e^{2}} \; \ll \sigma$. 
\subsection{Imperfect experiments}

For reasons beyond the observer's control, experiments in astronomy are often
imperfect, yielding data from which a definitive solution cannot be obtained. 
In double star astronomy, 
examples are long-period binaries that have only completed 
a fraction of an orbit since discovery. If a solution is nevertheless attempted,
orbits with very different parameters may provide acceptable fits - 
see Fig.2 in L14a and references therein.
Among these acceptable orbits may be orbits that violate the CPr, as is the 
case for the nearly parabolic orbit in that figure.

Comparison of the simulations here with those for incomplete orbits in L14a 
is illuminating. Here and in Pourbaix (2002), we find CPr violations  
even though the orbit is complete $(P_{*} < t_{M})$. This shows
that a weak orbital signal suffices for the experiment
to be imperfect and to thereby permit solutions that violate the CPr.
 
When the min-$\chi^{2}$ elements violate the CPr, we might suspect that there 
exists a better solution that, despite a higher $\chi^{2}$, should be preferred
because it is consistent with the CPr. 
\section{A Bayesian prior derived from the CPr}

The CPr is now treated as an integral part of
Bayesian estimation and not as 
an a posteriori arbiter of a solution's believability. This is achieved
by constructing a Copernican prior.
\subsection{Conventional priors}

If $H$ denotes the hypothesis and $D$ the data and if $I$ is some
relevant information, then by Bayes' theorem (Jaynes 2003, p.85), 
the posterior density of $H$ given $D$ and $I$ is 
\begin{equation}
  Pr(H|D,I) \: \propto  \: Pr(H|I) \: Pr(D| H,I) 
\end{equation}
Here $Pr(D| H,I) \equiv {\cal{L}}(H,I|D)$ is the 
likelihood, and $Pr(H|I)$ is the prior probability of $H$ given $I$. 

If there is no information $I$, the prior reduces to $Pr(H)$
and so becomes the {\em subjective} choice of the investigator.  
This aspect of Bayesian estimation is controversial and much-debated.
However, there is little reason to object to current astronomical practice  
with regard to $Pr(H)$ since the aim is 
not to quantify prejudice but to admit ignorance.
Thus {\em flat} priors are typically imposed on the 
parameters of $H$. Moreover, the ranges over which these priors are non-zero
are chosen to comfortably enclose the intervals within which there
is significant likelihood ${\cal{L}}$ and therefore significant
posterior density $Pr(H|D)$. Such priors are {\em non-informative}. 

In the {\em Gaia} problem, it is tempting to use a variant of this methodology
to eliminate CPr-violating orbits. Thus, Fig.1 
suggests a prior on $e$ that is zero for $ e > 0.6$.
But this would be an {\em ad hoc} fix for this particular data set.
A Bayesian prior should not depend on, nor be derived from, $D$. 
\subsection{A Copernican prior}

For the problem under consideration, the symbols $H,D$ and $I$ are
defined as follows:  \\

H: The components of the {\em theoretical} scan vector 
$\vec{s}$ are
$s_{n} = s(t_{n},\alpha_{n}; \vec{\theta})$, the 
predicted 
abscissae at the known times $t_{n}$ and scanning angles $\alpha_{n}$ for the 
Keplerian orbit $\vec{\theta}$. \\

D: The elements of the data vector $\tilde{\vec{s}}$ are 
$\tilde{s}_{n}$, the measured abscissae at $(t_{n},\alpha_{n})$. \\

I: Orbits $\vec{\theta}$ with random orientations 
and random shifts in epoch  
are all equally probable {\em a priori}  . \\       
\\
Comments: \\ 

(i) Information $I$ is the means of incorporating the CPr.\\ 

(ii) Since orbits are periodic,  
choosing a random value of $\tau = T/P \in (0,1)$ is equivalent to a random 
shift in epoch. \\ 

(iii) Imposing $I$ is appropriate for orbits 
{\em discovered} with {\em Gaia} but not if
a previously-known orbit is targeted.\\
\\
Given that the hypothesis of Keplerian motion enters via the theoretical
vector $\vec{s}$, the Copernican prior $Pr(H|I)$ becomes $\pi(\vec{s}|I)$, the 
probability density at $\vec{s}$ when $I$ is taken into account. However,
$I$ does not itself suffice to determine this probability density function
(PDF).
In addition, $P,e$ and $a$ - or their prior distributions - must be specified.
We choose the latter option on the grounds of simplicity.

For the bounded quantity $e$, we assume a uniform prior in $(0,1)$.
For the unbounded positive continuous parameters $P$ and $a$, it is 
appropriate (Jaynes 2003, p.395) to assign equal prior probabilities to equal 
logarthmic intervals - i.e., Jeffreys' priors.

With these additional assumptions, the Copernican prior is determinate
and given by
\begin{equation}
 Pr(H|I) \propto  a^{-1} \pi_{1}(\vec{s}|I)   
\end{equation}
where $\pi_{1}(\vec{s}|I)$ is the PDF at $\vec{s}$ for orbits scaled to
$a = 1 \arcsec$.

\subsection{Accurate treatment}

A rigorous calculation of $\pi_{1}(\vec{s}|I)$   
would proceed as follows: The scale parameter $a$ is set $= 1\arcsec$,
and $e$ and $\log P$ are randomly chosen in $(0,1)$ and 
$(\log P_{L}, \log P_{U})$, respectively.    
The orbit's orientation $(\rm i, \omega, \Omega)$ and 
epoch $\tau = T/P$
are randomly chosen, and the resulting theoretical 
$\vec{s}$ computed. These steps are repeated many 
times, thus 
generating points that
populate the $N$-dimensional $\vec{s}$-space with a probability density 
determined by $I$ and by the prior distributions of $e$ and $P$.
As the sample size $\rightarrow \infty$, the result is the 
desired PDF $\pi_{1}(\vec{s}| I)$. 
However, with $N \sim 70$, this brute-force approach is not 
feasible. A less rigorous approach must be adopted.

\subsection{Approximate treatment}

Consider an orbit with $a = 1\arcsec$ and eccentricity $e$.
Sampled with random orientations and epochs, the theoretical abscissae $s_{n}$ 
will extend over the full permitted range, namely
$-(1+e)$ to $+(1+e)$. 
In other words, whatever the values of $(t_{n},\alpha_{n})$,
there will be some combination of orientation $(\rm i, \omega, \Omega)$ and 
epoch $\tau$ for which these limits are reached. 
Given that $e \in (0,1)$, it follows that the $s_{n}$-values 
populate the interval $(-2,+2)$.    
Since this applies to every component of
$\vec{s}$, the distribution of the vectors $\vec{s}$ in N-dimensional space
is approximately isotropic. Accordingly, most of the information relevant to 
CPr violations is contained in the distribution 
of the Euclidean ``lengths'' of the vectors $\vec{s}$. We therefore define the 
statistic $\xi$ given by
\begin{equation}
  \xi^{2} \: = \: N^{-1} \sum_{n} (s_{n}/a)^{2}
\end{equation}

With this statistic as the sole basis for assessing CPr violations,
the approximate Copernican prior is
\begin{equation}
 Pr(\: H| \: I) \: \propto  \:  a^{-1} \: \pi_{1}(\: \xi|\: I)  
\end{equation}
where  $\pi_{1}(\: \xi|\: I)$ is the PDF of the lengths $\xi$ for orbits
scaled to $a = 1\arcsec$.        
This prior is used in the subsequent tests. 
\subsection{Calculation of $\pi_{1}(\: \xi| \: I)$ }

With the above assumptions, the problem has been reduced to
tabulating the 1D function $\pi_{1}(\xi|I)$. The steps  
are as follows:\\    

1) The campaign $(t_{n},\alpha_{n};N)$ is specified (Sect.2.3).\\

2) We set $a = 1\arcsec$ and choose random values of $ e \in (0,1)$ 
and $\log P \in (\log P_{L}, \log P_{U})$.\\

3) A random orientation is selected by taking 
$\omega= 2 \pi z_{u}, \: \Omega = \pi z_{u}$, and 
$cos \: \rm i = 1 - 2 z_{u}$.\\    

4) A random epoch is selected by taking $\tau = T/P = z_{u}$.\\

5) With the orbit vector $\vec{\theta}$ determined in steps 2)-4), the   
coordinates $(x_{n},y_{n})$ at $t_{n}$ are computed.\\

6) From these coordinates, the components of $\vec{s}$ are given by
Eq.(2), and the corresponding length $\xi$ by Eq.(10).\\   

7) Steps 2) - 6) are repeated $10^{8}$ times. The resulting histogram of
$\xi$-values gives $\pi_{1}(\xi|I)$.\\

Comments: \\  

(i) To eliminate parabolic orbits and to avoid convergence failures when solving
Kepler's equation, an upper limit $e = 0.999$ is imposed at step 2).\\
  
(ii) Since the exact period is known, we take $P_{L} = 1y$ and 
$P_{U} = 10y$. \\

(iii) From the $10^{8}$ values of $s_{n}$ for each $n$, the maximum and minimum
values are derived. These closely approach the expected values $\pm 2$   
(Sect. 4.4).\\

(iv) If no assumptions are made about the prior distributions of $P$ and $e$,
then $\pi_{1} = \pi_{1}(\xi;P,e|I)$, thus requiring an extra two dimensions in 
its tabulation. This is feasible, but the preference 
here is to investigate the simplest formulation.\\

The resulting accurate determination of $\pi_{1}(\xi|I)$ is plotted in Fig.3.
This shows that the astrometric lengths are typically in the interval
$(0.4,1.1)$, and that values $\la 0.3$ are improbable.

\begin{figure}
\vspace{8.2cm}
\includegraphics{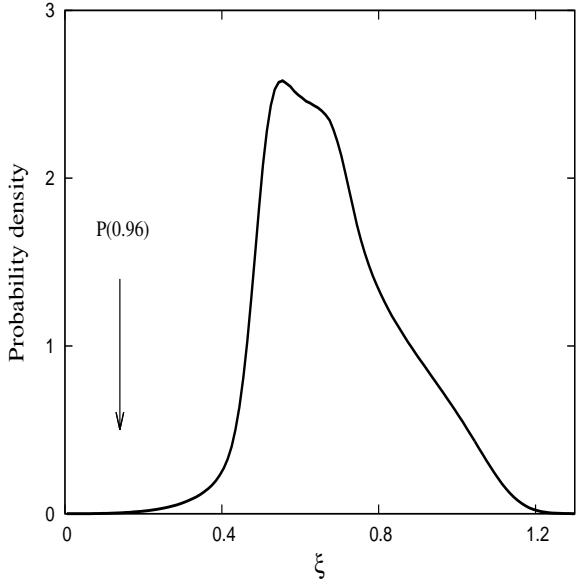}
\caption{The PDF $\pi_{1}(\xi|I)$ for astrometric ``lengths'' $\xi$ defined by
Eq.(10).
The orbits have random
orientations $(\rm i, \omega, \Omega)$, random epochs $\tau$,
random eccentricities $\in (0,1)$, and random values of 
$\log P \in (0.0,1.0)$.  
The corresponding PDF for P-orbits $(\rm i = \omega = 90\degr)$ 
with $e=0.96$ is a 
near delta function at $\xi = 0.14$.} 
\end{figure}
The position of a P-orbit in this plot is of interest. The above steps 
are therefore repeated
with the constraints $e = 0.96$ and $\rm i = \omega = 90 \degr$. For this
orbit, the range for the $s_{n}$ is $(-b,+b)$
or $(-0.28,+0.28)$ in units of $a$. 
Consistent with this, the PDF is a near delta function at $\xi = 0.14$, 
and this 
location is indicated in Fig.3. The probability of obtaining an even smaller
value is 
\begin{equation}
 \Pi_{1} (\xi| I) = \int_0^{\xi} \pi_{1}(\xi|I) \: d \xi 
\end{equation}
which gives $\Pi_{1}(0.14|I) = 1.7 \times 10^{-4}$, showing 
that P-orbits populate an extremely low probability tail of $\pi_{1}(\xi|I)$. 
\section{Bayesian estimation subject to the CPr}

Formulae are now developed that allow the Copernican 
prior to be included in the calculation of posterior densities and
credibility intervals.
\subsection{Posterior densities}

For every orbit in $\vec{\theta}$-space, there is a theoretical scan vector
$\vec{s}$ corresponding to the scanning campaign 
$(t_{n}, \alpha_{n})$.
From this $\vec{s}$ and the orbit's $a$, we can compute $\xi$ from Eq.(10).
Then, from $\xi$, we obtain $\pi_{1}(\: \xi | \: I)$ 
by interpolating in the data file plotted in Fig.3. This procedure results in 
an ensemble of orbit vectors weighted according to their Copernican priors 
$a^{-1} \pi_{1}(\: \xi | \: I)$, and so CPr violations are penalized.
From this ensemble, the Bayesian machinery then
computes posterior densities by 
further weighting the orbits in accordance with their goodness-of-fits
to the {\em measured} scan vector $\tilde{\vec{s}}$.

The posterior density at $(\vec{\phi}, \vec{\psi})$ is 
\begin{equation}
  \Lambda(\vec{\phi}, \vec{\psi} | D,I)   \propto a^{-1} \: \pi_{1}(\xi|I) 
                             \:  {\cal L} (\vec{\phi}, \vec{\psi}| D) 
\end{equation}
Ignoring coefficients independent of $(\vec{\phi}, \vec{\psi})$ and
assuming normally distributed measurement errors, we have
\begin{equation}
 {\cal L} \: \propto  \: \exp ( -\frac{1}{2} \: \hat{\chi}^{2}) \times
                        \exp ( -\frac{1}{2} \: \delta \chi^{2})
\end{equation}
where   
$\hat{\chi}^{2}(\vec{\phi})$ is the minimum at $\hat{\vec{\psi}}$, and   
$\delta  \chi^{2}$ is the postive increment due to the displacement
$\vec{\psi} - \hat{\vec{\psi}}$ at fixed $\vec{\phi}$.

In the absence of the Copernican prior, $\Lambda \propto {\cal L}
\propto Pr(\phi) \times Pr(\psi| \phi)$. The first of these PDFs
is sampled at grid points $(i,j,k)$ giving weight factors
$ \propto exp(-\chi^{2}_{ijk}/2)$. 
The second PDF is randomly sampled as
described in Appendix A.4. If ${\cal N}_{ijk}$ is the number of
random points $\psi_{\ell}$ selected in $\psi$-space at $\phi_{ijk}$,
then each has weight  ${\cal N}_{ijk}^{-1}$.

With the Copernican prior included, the PDF $\Lambda$ given in Eq.(13)
is represented by a cloud of discrete orbit vectors   
\begin{equation}
\vec{\theta}_{m} \equiv (\vec{\phi}_{ijk}, \vec{\psi}_{\ell})  
\end{equation}
with weights
\begin{equation}
  \mu_{m} = a_{m}^{-1} \pi_{1} (\xi_{m}|I)  \times 
                      \: {\cal N}_{ijk}^{-1}
          \:   \exp ( -\frac{1}{2} \: \hat{\chi}^{2}_{ijk})
\end{equation}
Here $m$ enumerates the random points  $\vec{\psi}_{\ell}$ across all
grid cells $(i,j,k)$.

From these weighted orbits, the posterior mean of a quantity  
$Q(\vec{\theta})$ given $D$ {\em and} $I$ is
\begin{equation}
 < Q >  \: = \: \sum_{m} \: \mu_{m} Q_{m} \: / \: \sum_{m} \: \mu_{m}
\end{equation}
and credibility intervals 
are derived as described in Sect.4.2 of L14b. 

This discrete representation of $\Lambda$ and the resulting credibility 
means and intervals become exact as the grid steps
$\rightarrow 0$ and the ${\cal N}_{ijk} \rightarrow \infty$.

\section{Numerical experiments}

The approximate theory (Sect.4.4) of the Copernican prior 
is now applied to the model binary defined in Eq.(1).
\subsection{Code verification}

In the strong-orbit limit, violations of the CPr are not an issue and
so, even with the inclusion of the Copernican prior, the posterior means 
should $\rightarrow$ the exact elements given in Eq.(1). To test this,
the Bayesian code is used to compute the solution
when $\log \beta = 1.5$. The posterior means and {\em equal-tail} 
$1\sigma$ credibility intervals for the elements are as follows: 
\begin{eqnarray}
  \log P(y) = 0.4625^{+0.0011}_{-0.0013} &      
    \;\;  e = 0.0529^{+0.0039}_{-0.0039}    \nonumber             \\ 
                &\tau = 0.384^{+0.012}_{-0.012}    \nonumber \\ 
  \log a/\sigma = 1.4990^{+0.0033}_{-0.0038} &    
  \rm i = 40\fdg6^{+1\fdg0}_{-1\fdg0}     \nonumber \\ 
  \omega = 144\fdg8^{+4\fdg9}_{-4\fdg9}   &     
                   \Omega = 69\fdg6^{+1\fdg3}_{-1\fdg3}   
 \end{eqnarray}
These results are consistent with expectation:
five of the seven credibility intervals include the exact values. Minor
deviations occur for $\tau$ and $\omega$.  

Note that the credibility interval for $\omega$ is substantially larger
than those for $\rm i$ and $\Omega$. This is a consequence of the small
eccentricity, since $\omega$ becomes indeterminate as 
$e \rightarrow 0$.

In this strong-orbit regime,
${\cal L}$ is sharply peaked in parameter space; consequently, $\xi$ and 
therefore $\pi_{1}(\xi|I)$ vary little within the narrow domain of high 
likelihood. It follows that posterior densities are then largely
determined by ${\cal L}$, which overwhelms the prior.
\subsection{Varying $\beta$}

Solutions are computed with $log \beta = -0.6 \:(0.05)\:1.2$, 
spanning the range from weak to strong 
orbits. For each $\beta$, the elements' posterior 
means and $1\sigma$ credibility intervals 
are derived as in Sect.6.1. In addition,
for each data vector $\tilde{\vec{s}}$, the min-$\chi^{2}$ solution
is computed as in Pourbaix(2002). Note that when $\beta$ changes, so does the 
random number seed.

In Figs. 4-6, the solution sequences are plotted  
for $\log a/\sigma$, $\: e$, and $\rm i$.
For $\log \beta \ga 0.2$, 
the credibility intervals are consistent with both the min-$\chi^{2}$ values
and with the exact values. However, at $\log \beta = -0.05$, major 
disagreements occur. The min-$\chi^{2}$ value of ${\log a/\sigma}$ 
suddenly jumps to $1.14$, which is 1.19 dex greater than the exact value.
Correspondingly, $e$ jumps to 0.9975 (the highest value allowed by the
grid) and $\rm i$ jumps to $89\fdg6$. Thus, at $\beta = -0.05$ dex, 
the min-$\chi^{2}$ solution is a P-orbit (Sects.3.2-3.4)

At this same $\beta = -0.05$ dex, the 
$1\sigma$ credibility intervals are 
$(-0.20,0.05)$ for $\log a/\sigma$, $(0.06,0.44)$ for $e$, and 
$(52\degr,83\degr)$ for $\rm i$. Thus, 
the Bayesian solution with Copernican prior does not undergo a  
transition into a P-orbit. In fact, the solution remains
(marginally) consistent with the exact solution.

For  $\log \beta \leq -0.05$, most of the min-$\chi^{2}$ solutions
are P-orbits.
But the Bayesian solutions with Copernican prior do not exhibit such
strikingly anomalies. Nevertheless, they 
do eventually $(\log \beta \la -0.3)$ become inconsistent with the exact
parameters.
\begin{figure}
\vspace{8.2cm}
\includegraphics{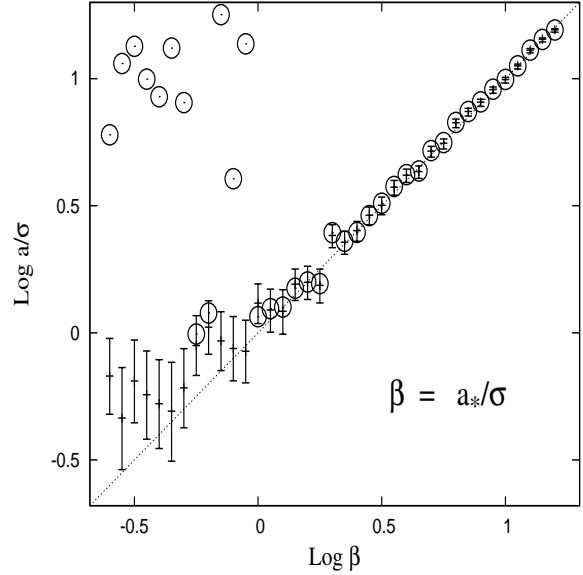}
\caption{Sequence of solutions for $\log a/\sigma$. 
The points with error bars are
the posterior means $<\! \log a/\sigma \!> $ plotted with 1-$\sigma$ credibility 
intervals.
The open circles are the min-$\chi^{2}$ values. The dotted line is
the locus of exact values $\log a_{*}/\sigma$.} 
\end{figure}
\begin{figure}
\vspace{8.2cm}
\includegraphics{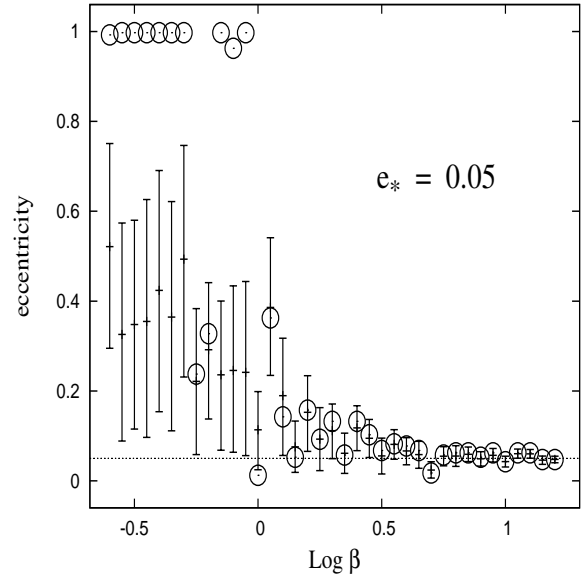}
\caption{Sequence of solutions for $e$.
The points with error bars are
the posterior means $<\! e \!>$ plotted with 1-$\sigma$ credibility intervals.
The open circles are the min-$\chi^{2}$ values. The exact value 
is $e_{*} = 0.05$.} 
\end{figure}
\begin{figure}
\vspace{8.2cm}
\includegraphics{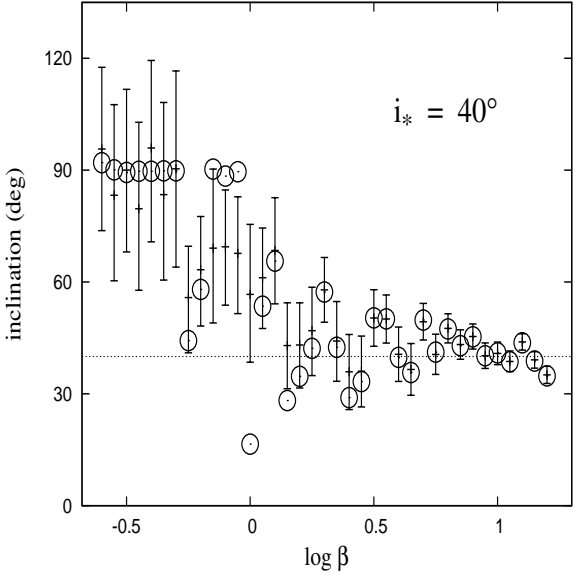}
\caption{Sequence of solutions for $\rm i$. The points with error bars are
the posterior means $<\! \rm i \!>$ plotted with 1-$\sigma$ credibility 
intervals. The open circles are the min-$\chi^{2}$ values. Orbits with
$\rm i > 90\degr$ are retrograde.
The exact value is $\rm i_{*} = 40\degr$.} 
\end{figure}
\subsection{Orbits from noise}

Figs. 4-6 show that even for extremely weak orbits
the Copernican prior has 
eliminated P-orbits. However,
the plotted credibility intervals reveal that when $\log \beta \la -0.3$
the posterior PDFs are 
systematically displaced from the exact values. 
This could indicate that the orbital signal is then too weak for detection,
a conclusion strongly supported by Fig.6 which shows that the posterior
densities of retrograde and 
prograde orbits are then about equal.

To investigate this issue further, the code is now used
to compute solutions when $a_{*} = \beta \sigma = 0$. 
The posterior density of 
$a/\sigma$ is plotted in Fig.7 for a particular 
realization of the noise vector $\tilde{\vec{s}}$.
For comparison, plots
with  $\beta = 0.5$ and $1.0$ are also included.

In 20 independent repetitions with $\beta = 0$, 
the range found for $<\! a/\sigma \!>$ is
0.45 to 0.81, with average $<<\!a/\sigma\!>> = 0.57$. For
$<e>$ the range is 0.21 to 0.53, with $<<e>> = 0.37$; and for $<\rm i>$ 
the range is
$67\degr$ to $113\degr$, with $<< \rm i >> = 88\degr$. 
Since these are
consistent with Fig.4-6 when $\log \beta \la -0.3$, we conclude that the
aforementioned
systematic displacements simply reflect the code's reponse to data 
with negligible orbital signal.

The bias in $a/\sigma$ for $\beta = 0$ evident in Fig.7 is reminiscent of the
bias in the eccentricities
of spectroscopic binaries for nearly circular orbits (Lucy \& Sweeney 1971).
In both cases, bias is the inevitable result of estimating a {\em non-negative}
parameter at or near its zero lower bound. From the values of
$<< \! a/\sigma \!>>$ plotted in Fig.7, the bias of
$a/\sigma$ is typically 0.57, 0.21 and 0.13 at $\beta = 0.0, 0.5$ and $1.0$,
respectively.

\begin{figure}
\vspace{8.2cm}
\includegraphics{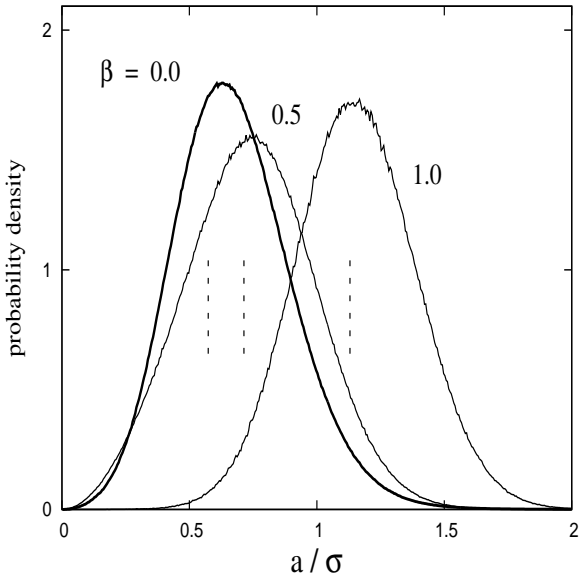}
\caption{The posterior densities of $a/\sigma$ for $\beta = 0.0, 0.5$ and 
$1.0$  
for particular realizations of the measurement vectors $\tilde{\vec{s}}$. 
The vertical dotted lines indicate the average positions
of the corresponding posterior means $<a/\sigma>$. Each of these is obtained
from 20 independent simulations.} 
\end{figure}

Also of interest when $\beta = 0$ is the frequency of P-orbits
for min-$\chi^{2}$ solutions. From 200 independent simulations, 156 or
$78\%$ are P-orbits - i.e., have $e > 0.95, \rm i \approx 90\degr$ and
$\omega \approx 90$ or $270\degr$. Thus, the PDFs of $e, \rm i$ and 
$\omega$ for min-$\chi^{2}$ solutions
when $\beta = 0$ are dominated by near delta functions at the
Pourbaix loci. 

Figs. 8-10 plot the posterior densities
of $e, cos \rm i$ and $\omega$ for $\beta = 0$ when the Copernican prior  
is included. While these plots are free from peaks at the Pourbaix loci,
they do show evidence of imperfections that presumably derive from
the approximate treatment of the Copernican prior (Sect.4.4). Ideally,
when analysing pure noise, the inferred values of $cos i$ and $\omega$ 
should be uniformly distributed in $(-1,+1)$ and $(0,360\degr)$,
respectively. Figs. 9-10 show departures from this ideal.

As in Fig.7, Figs.8-10 also include solutions for $\beta = 1$. Figs.8 and 9
show the emergence of peaks at the exact values of $e$ and $cos \rm i$,
repectively. However, an emerging peak is not evident at $\omega_{*}$
in Fig.10. This is due to the near indeterminacy of $\omega$ when
$e \ll 1$ - see Sect.6.1.

\begin{figure}
\vspace{8.2cm}
\includegraphics{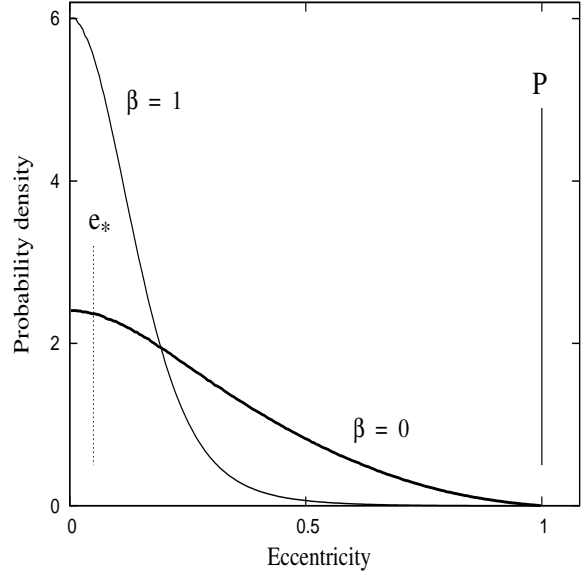}
\caption{Posterior PDFs for $e$ when $\beta = 0$ and $1$. The Pourbaix peak
at $e = 1$ and the exact value $e_{*} = 0.05$ are indicated.} 
\end{figure}
\begin{figure}
\vspace{8.2cm}
\includegraphics{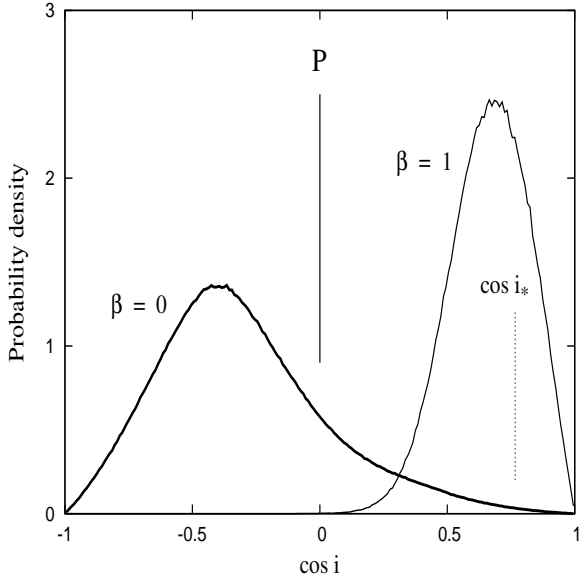}
\caption{Posterior PDFs for $\cos \rm i$ when $\beta = 0$ and $1$. The Pourbaix
peak at $cos \rm i = 0$ and the exact value $cos \rm i_{*}$ are
indicated.} 
\end{figure}
\begin{figure}
\vspace{8.2cm}
\includegraphics{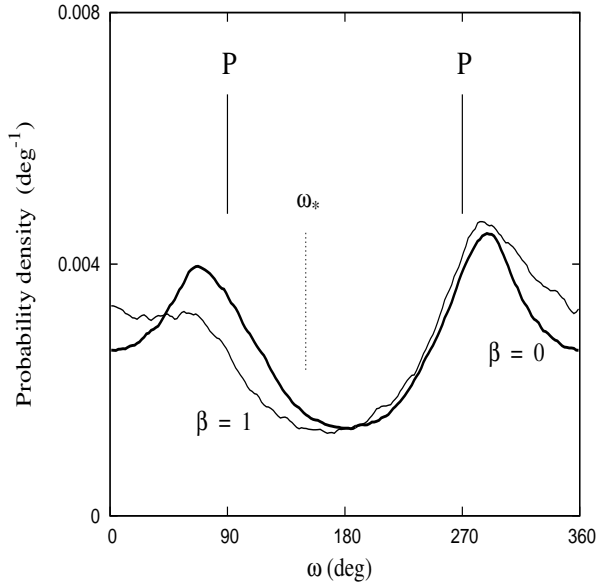}
\caption{Posterior PDFs for $\omega$ when $\beta = 0$ and $1$.
The Pourbaix peaks at $\omega = 90, 270\degr$ and the exact value 
$\omega_{*} = 150\degr$ are indicated.} 
\end{figure}
\subsection{$\xi$-probabilities}

Given that P-orbits arise when applying a conventional data analysis
technique to synthetic {\em Gaia} data, there is some danger that such orbits
will contaminate the huge data bases expected from the
{\em Gaia} mission. On the assumption that this Bayesian procedure
cannot feasibly replace the existing
pipeline analyses, a less ambitious approach to elimating
P-orbits is desirable. 
 
Let $\vec{\theta}_{0}$ be the min-$\chi^{2}$ elements
derived from an observed scan vector $\vec{\tilde{s}}$ and let the
corresponding fitted vector be $\vec{s}_{0}$. From $\vec{s}_{0}$,
the astrometric length $\xi_{0}$ of the orbit $\vec{\theta}_{0}$
is then given by Eq.(10).
This scale-free length refers to an orbit with physical
parameters $(P_{0}, e_{0})$ observed at epoch $\tau_{0}$ and
orientation $(\rm i_{0}, \omega_{0}, \Omega_{0})$. We now define $p_{0}$
to be the probability that a shorter length $\xi$ would be found with random
epochs and orientations but with $P$ and $e$ fixed at their   
min-$\chi^{2}$ values. Thus, with steps 3) -6) of Sect.4.5, we compute
\begin{equation}
 p_{0} = Pr(\xi < \xi_{0} | P_{0}, e_{0})
\end{equation}
\begin{figure}
\vspace{8.2cm}
\includegraphics{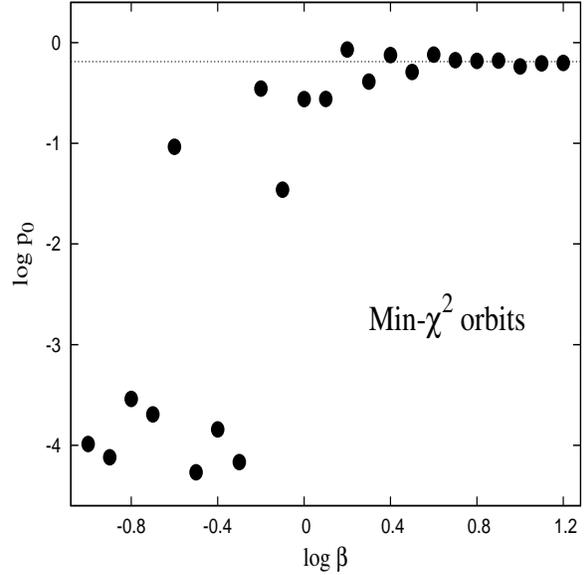}
\caption{$\xi$-probabilities $p_{0}$ for a $\beta$-sequence of min-$\chi^{2}$ 
orbits computed for simulations of the orbit defined by Eq.(1). The
dotted line is the exact value $p_{*} = 0.649$.} 
\end{figure}

In Fig.11, $\log p_{0}$ is plotted against $\log \beta$ 
for min-$\chi^{2}$ orbits. For strong orbits, the values scatter about the 
exact value $= -0.188$. But for weak orbits the solutions are 
the CPr-violating P-orbits with $\log p_{0} \sim -4$.
Accordingly, if a {\em Gaia} orbit catalogue were contaminated by P-orbits
a cut excluding orbits with $p_{0} < -3$ dex would remove them.

Besides P-orbits, other as yet unrecognized anomalies, biases and 
selection 
effects may be present in a {\em Gaia} catalogue. Accordingly, it is worth noting 
that there are {\em five} 
quantities which, in a {\em perfect catalogue}, are uniformly and independently
distributed in $(0,1)$. These quantities are: $(1+cos \rm i)/2,
\: \omega/2 \pi, \: 
\Omega/\pi, \: \tau$ and $p_{0}$. This statement yields {\em fifteen}
statistical tests that should be applied to a catalogue of {\em Gaia} orbits. 
\subsection{Detecting a second companion}

A further weak-orbit problem for {\em Gaia} is that of detecting a second
companion ($B$) when the first ($A$) is well-determined. This is a
goodness-of-fit problem: the presence of $B$ degrades the fit achieved
when only $A$ is considered.    

To investigate this problem, {\em Gaia} data is created (Sects.2.3,2.4) for a star
with invisible companions $A$ and $B$. Companion $A$ has the elements given
in Eq.(1) with $\beta_{A} = 10$, and $B$ is in a coplanar orbit with  
$P = 7.2y, e = 0.2, \tau = 0.7$, and a reflex orbit with 
semi-major axis $= \beta_{B} \sigma$. 

A 1-D sequence of {\em Gaia} scans is created for this two-companion model
with $\log \beta_{B} = -0.6 (0.05) 0.6$, and each scan is analysed with the
Bayesian code under the assumption of only {\em one} companion. 

For each $\beta_{B}$, the code creates (Sect.5) a cloud of orbits 
$\vec{\theta}_{m}$ with weights $\mu_{m}$. 
The $\chi^{2}$ of the $m$-th orbit's fit to the data vector 
$\tilde{\vec{s}}$ is 
\begin{equation}
 \chi^{2}_{m} = \hat{\chi}^{2}_{ijk} + \delta \chi^{2}_{\ell}
\end{equation}
Now, if the one-companion solution provides a satisfactory fit, then
orbits of high weight should have $\chi^{2}_{m} \la N$. On the 
other hand, if the solution is not satisfactory, then these high-weight
orbits 
will have 
$\chi^{2}_{m} > \chi^{2}_{N,\alpha}$ with $\alpha < 0.05$. 
These expectations can be reduced to a single measure of
goodness-of-fit, namely $<\! \chi^{2} \!>$, the posterior mean of $\chi^{2}$.
This is calculated from Eq.(17) with
$Q_{m} =  \chi^{2}_{m}$. 

The values of $<\! \chi^{2} \!>$ are plotted against $\log \beta_{B}$ in Fig.12.
As always with statistical tests, the investigator has discretion as to 
when he deems a model to be successful. 
In this case, he is likely to suspect an additional orbit
when  $\log \beta_{B} \geq 0.3$.
On the other hand, scans with  $\log \beta_{B} \leq 0.1$ are fitted
with $<\chi^{2}>  \: \approx \: \tilde{\chi}^{2}$, the residuals are
therefore consistent with measurement errors and so there is no 
evidence of an additional orbit.
\begin{figure}
\vspace{8.2cm}
\includegraphics{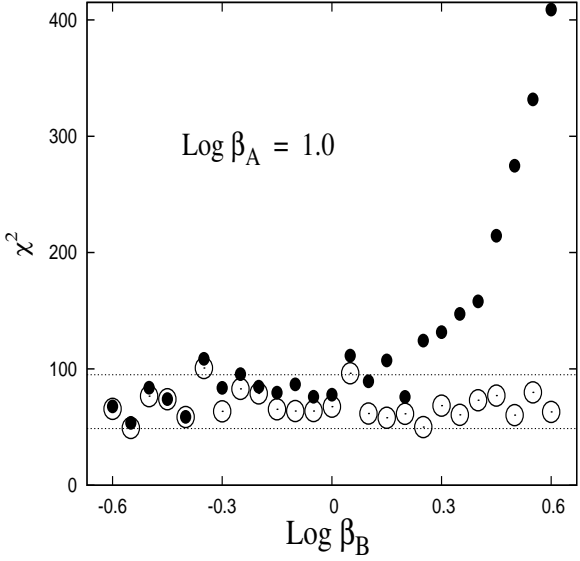}
\caption{Detecting a second companion. The filled circles are the
posterior means $<\chi^{2}>$ measuring the goodness-of-fits of the Bayesian 
{\em single-orbit} solutions to the simulated scan vectors $\tilde{s}$ for the
{\em two-orbit} model. The amplitudes are $\beta_{A} = 10$ and 
$\log \beta_{B} = -0.6(0.05)0.6$. The open circles are the corresponding
values of $\tilde{\chi}^{2}$ given by Eq.(4). The dotted lines are
the $\chi^{2}_{N,\alpha}$ values for $\alpha = 0.025, 0.975$ with
$N = 70$.} 
\end{figure}
\section{Conclusion}

The aims of this paper are twofold. First, to provide a weak-orbit
analysis for {\em Gaia} and, in particular, to investigate the occurrence
of the spurious solutions found by Pourbaix (2002).
Secondly, to use the {\em Gaia} problem as a test case for 
a procedure that incorporates
the CPr into the machinery of statistical astronomy.

With regard to spurious solutions, Pourbaix's (2002) finding is
confirmed and the puzzle of nearly parabolic, edge-on orbits explained 
in terms of the near degeneracy (Sect.3.4) of scan vectors $\vec{s}$ under
the addition of such orbits. Moreover, these orbits are shown to
violate the CPr (Sect.3.3) and do not arise when the CPr is adopted
as a fundamental postulate in Bayesian estimation (Sects.5 and 6). 

More generally, incorporating the CPr in statistical analyses may 
improve solutions derived for imperfect experiments (Sect.3.5).
In addition to poor precision and limited sampling, weather, the seasons, 
atmospheric opacity and interstellar extinction 
are among the numerous factors that result in data sets that are less 
than ideal.

When an astronomer must perforce analyse an imperfect data set, he needs
to be aware that supposedly optimum statistical procedures - e.g.,
min-$\chi^{2}$ or Bayesian estimation with non-informative priors -
can in extreme cases, as with the P-orbits, give anamolous
solutions. 
Moreover, at a more subtle level, even when an anomaly is not
immediately evident, the complicated topology (Sect.3.2) of the
likelihood function implies
that the above 'optimum' procedures are unlikely to be so. 
An investigation of
how an estimation procedure can exploit imperfect data 
should be carried out (Sect.3.2) and an appropriate prior constructed. Often 
Copernican considerations with regard to position, epoch or orientation will
be crucial.            

\acknowledgements

I thank A.H.Jaffe and D.J.Mortlock for helpful discussions on Bayesian
methods and the referee for justified criticisms of the original version.

\appendix

\section{Statistics in Thiele-Innes space}

In L14a,b, each observation of the model visual 
binary yielded {\em two} measurements $(\tilde{x}_{n},\tilde{y}_{n})$,
the sky coordinates of the secondary's displacement from the
primary at time $t_{n}$. In this circumstance, minimizing $\chi^{2}$
to obtain the Thiele-Innes constants $\hat{\psi}_{j}$ separates into two 
independent problems, minimizing the $x-$coordinate contribution to 
$\chi^{2}$ to obtain $(A,F)$ and minimizing the $y-$coordinate contribution to
obtain $(B,G)$. This separation results in the considerable simplifications
exploited in L14a,b. 

However, these simplifications are lost when observing
an astrometric binary with a 1-D scanning device.
On the assumption that the parallactic and proper motion have been subtracted,
the measurement at $t_{n}$ with scanning angle $\alpha_{n}$  
is the observed star's displacement $\tilde{s}_{n}$ from the binary's 
barycentre.
\subsection{Normal equations}
    
For given orbit 
$\vec{\theta} \equiv (\vec{\phi},\vec{\psi})$, the goodness-of-fit
criterion $\chi^{2}(\vec{\phi},\vec{\psi})$ is given by Eq.(5).
At fixed $\vec{\phi}$, the orbit $(x,y)$ is linear in $\vec{\psi}$. 
Accordingly,
since $s_{n} = 0$ when $\vec{\psi} = \vec{0}$, the predicted abscissa at 
$t_{n}$ is
\begin{equation}
 s_{n} = \sum_{j}  \left( \frac{\partial s}{\partial \psi_{j}} \right)_{n}
                              \: \psi_{j} 
\end{equation}
Substitution of $s_{n}$ into Eq.(5) then allows    
the min-$\chi^{2}$ solution for the Thiele-Innes
vector $\vec{\psi}$ to be obtained without iteration.
The normal equations are 
\begin{equation}
   {\cal A}_{ij} \psi_j = b_{i}
\end{equation}
where, the curvature matrix,
\begin{equation}
   {\cal A}_{ij} = \frac{1}{\sigma^{2}} \sum_{n} 
   \left( \frac{\partial s}{\partial \psi_{i}}
 \right)_{n}  \left( \frac{\partial s}{\partial  \psi_{j}} \right)_{n}
\end{equation}
and  
\begin{equation}
    b_{i} = \frac{1}{\sigma^{2}} \sum_{n} \tilde{s}_{n} 
  \left( \frac{\partial s}{\partial \psi_{i}} \right)_{n}   
\end{equation}
The partial derivatives in these equations can be expressed in terms
of the elliptical rectangular coordinates $X(E),Y(E)$ via Eq.(2) and
Eq.(A.2) of L14a. 

The solution of Eq.(A.2) is 
$\hat{\vec{\psi}} = (\hat{A},\hat{B},\hat{F},\hat{G})$ and we write
$\hat{\chi}^{2}(\vec{\phi})  = \chi^{2}(\vec{\phi},\hat{\vec{\psi}})$

\subsection{Increment in $\chi^{2}$}

At fixed $\vec{\phi}$, a displacement $\delta \vec{\psi}$ from
$\hat{\vec{\psi}}$ results in a positive increment $\delta \chi^{2}$.
The abscissa corresponding to this displacement is given by Eq.(A.1).
Substitution in Eq. (5) then gives
$\chi^{2} = \hat{\chi}^{2} + \delta \chi^{2}$. From the quadratic terms
in the resulting expression, we obtain 
\begin{equation}
  \delta \chi^{2} = \delta \vec {\psi}^{'} \vec{ {\cal A} } \: 
                                        \delta \vec {\psi}  
\end{equation}
\subsection{Probability density function $p(\vec{\psi}|\vec{\phi},D)$}

The distribution of probability at fixed $\vec{\phi}$ is a quadrivariate
normal distribution centred on $\hat{\vec{\psi}}$. 
If $\vec{\Sigma}$ is the covariance matrix, 
then 
\begin{equation}
 p = \frac{1}{4 \pi^{2}} \frac{1}{ \sqrt{| \vec{\Sigma} |}} \:
   \exp(- \frac{1}{2} \delta \vec {\psi}^{'} \vec{\Sigma}^{-1} 
                                        \delta \vec {\psi}  )
\end{equation}   
(James 2006, p.67).
Since $\vec{\Sigma}^{-1}  = \vec{ {\cal A}}$, comparison with Eq.(A.6) gives 
\begin{equation}
 p = \frac{1}{4 \pi^{2}} \frac{1}{ \sqrt{| \vec{\Sigma} |}} \:
   \exp(- \frac{1}{2} \delta \chi^{2})
\end{equation}

\subsection{Random sampling in $\psi$-space}

A random point $\delta \vec{\psi}_{\ell}$ sampling  
$p(\vec{\psi}|\vec{\phi},D)$ is obtained as follows 
(Gentle 2009, pp. 315-316): The first step is to compute the
Cholesky decomposition (Press et al. 1992, pp.89-91) of 
$\vec{\Sigma}$. Thus, we find  
the lower triangular matrix $\vec {L}$ such that  
\begin{equation}
      \vec {L} \vec {L}^{'} = \vec{\Sigma}
\end{equation}   
Now let $\vec {z}_{G}$ be a 4-D vector
whose elements are 
independent random Gaussian variates sampling ${\cal N}(0,1)$. Then 
\begin{equation}
      \delta \vec{\psi} =   \vec {L} \vec {z}_{G}  
\end{equation}
is a random displacement from $\hat{\vec{\psi}}$ satisfying the PDF
given by Eq.(A.7) 

If we generate ${\cal N}$ independent displacements, then the points
$\vec{\psi}_{\ell} = \hat{\vec{\psi}} + \delta \vec{\psi}_{\ell}$ 
give us the approximation
\begin{equation}
     p(\vec{\psi}|\vec{\phi},D)  = {\cal N}^{-1} \sum_{\ell} 
                  \delta(\vec{\psi} - \vec{\psi}_{\ell}) 
\end{equation}   
which is exact in the limit ${\cal N} \rightarrow \infty$.

\end{document}